\documentclass[a4paper]{article}
\usepackage{graphicx}

\begin{document}

\begin{flushright} {\large SAGA-HE-153-99} \end{flushright}

\vspace{1cm}

\begin{center}
{\LARGE Effective Vector Meson Masses in a Cutoff Field Theory}
\end{center}

\vspace{1cm}

\centerline{Katsuaki Sakamoto\footnote{E-mail address: 98td22@edu.cc.saga-u.ac.jp},
Manabu Nakai, Hiroaki Kouno, Akira Hasegawa and Masahiro Nakano$^*$}

\centerline{Department of Physics, Saga University, Saga 840, Japan}

\centerline{*University of Occupational and Environmental Health, Kitakyushu 807, Japan}

\vspace{5mm}

\centerline{\large \bf Abstract}
Based on quantum hadrodynamics with a finite cutoff, the effective masses of vector mesons($\omega$, $\rho$) in nuclear medium are calculated. 
We use a low-energy effective Lagrangian which is obtained by integrating high-energy quantum fluctuations. 
Although we use an artificial cutoff, the cutoff-dependence can be removed order by order.  
It is shown that there is a strong correlation between the effective $\omega$
-meson mass and the effective nucleon mass at the normal density. 
It is also found that the effective $\rho$-meson mass $m_\rho^*$ decreases as density increases. 
The rate of the decrease becomes smaller at high density. 
As a result, at the normal density, the $m^*_\rho/m_\rho$ is 0.85$\sim$0.95.

\vspace{5mm}

\section{Introduction}

$\quad \,$ The effective meson mass in nuclear medium is one of the most interesting issues in intermediate energy physics. 
Much attention has been paid to predicting the meson masses in nuclear medium
from various approaches such as quantum hadrodynamics (QHD) 
\cite{rf:Jean}-\cite{rf:Nakano4},
the QCD sum rules \cite{rf:Hatsuda,rf:Asakawa} and the quark meson coupling model \cite{rf:Saito}.  

The QHD has achieved considerable success in describing the bulk properties of nuclei. \cite{rf:Walecka,rf:Serot} 
The model has been used to discuss such diverse topics as collective modes 
in nuclear medium \cite{rf:Lim}, saturation problems
\cite{rf:Nakano1}-\cite{rf:Nakano3},
isoscalar magnetic moments \cite{rf:McNeil,rf:Furnstahl2}, electroweak \cite{rf:Wehrberger,rf:Horowitz,rf:Kurasawa2}, hadronic responses \cite{rf:Kormanyos}, and superfluidity in nuclear matter \cite{rf:Matsuzaki} with considerable success. 
Naturally the model is used for calculating the effective meson masses in nuclear medium \cite{rf:Jean}-\cite{rf:Nakano4}.
 
One important problem of calculation of
the meson mass is to estimate the quantum vacuum fluctuation effects.
As is in the case of QED, ultraviolet divergences appear in QHD, when we calculate vacuum fluctuation effects. 
Chin \cite{rf:Chin} estimated the vacuum fluctuation effects in the Hartree approximation by using a renormalization procedure, and found that vacuum fluctuation effects make the incompressibility of nuclear matter smaller and closer to the empirical value than in the original Walecka model. 
The renormalization procedure is also used to remove the divergences which appear in calculating meson self-energy. \cite{rf:Furn}\cite{rf:Jean}-\cite{rf:Nakano4} 

The renormalization is a well established procedure. 
However, it is ordinarily used in the theory which is applicable at any energy scale. 
On the other hand, it is natural to consider that QHD is a low-energy effective theory of QCD and is not valid at very high energy. 
In this point of view, a finite cutoff or a form factor should be introduced into the theory of QHD. 
One may introduce the cutoff \cite{rf:Kohmura} or the form factor \cite{rf:Furn} to avoid the instability of the meson propagators in the random phase approximation (RPA).  
Cohen \cite{rf:Cohen} introduced a four-dimensional cutoff into the relativistic Hartree calculations and found that the vacuum energy contribution may be somewhat different from that in the ordinary renormalization procedure, if the cutoff is not so large. 

One troublesome problem in use of the finite cutoff is that physical results depend on a value of the cutoff and a form of the regulator which is introduced into the theory by hand. \cite{rf:Nakano4}
It is difficult to determine a suitable value of cutoff and an appropriate form of the regulator phenomenologically. 
To overcome this difficulty, we use the low-energy effective Lagrangian which is used in the non-perturbative renormalization group (NPRG) method. \cite{rf:Wilson,rf:Polch} 
This effective Lagrangian can be defined at any artificial cutoff $\Lambda$, which is smaller than a limitation energy scale $\Lambda_0$ of the theory, and is obtained by integrating high-energy quantum fluctuations from $\Lambda_0$ to $\Lambda$. 
If we determine the couplings in this effective Lagrgangian phenomenologically,  we can calculate any physical quantities below the energy scale $\Lambda$. 
We do not need the information of the physical cutoff $\Lambda_0$. 
The phenomenological determinations of the couplings in the effective Lagrangian also remove the $\Lambda$-dependence of the physical results. \cite{rf:Lepage} 
In the series of papers \cite{rf:Kouno1,rf:Kouno2,rf:Kouno3}, we have studied nuclear medium properties and the vertex corrections in the framework of QHD with a finite cutoff, eliminating the cutoff dependence by determining the couplings in the low-energy effective Lagrangian.  
Based on these results, in this paper, we calculate the effective vector meson masses in the cutoff field theory (CFT). 

This paper is organized as follows. 
In {\S}2, we review the low-energy effective Lagrangian in the Hartree approximation. 
In {\S}3, we formulate the renormalization for the vector-meson self-energy in the RPA calculations with a finite cutoff. 
Using the formalism, we study the effective vector meson masses in nuclear medium in {\S}4. 
Section 5 is devoted to summary.

\section{Effective Lagrangian in the Hartree approximation}

$\quad \,$ Referring to our previous work \cite{rf:Kouno1,rf:Kouno2}, 
we introduce a low-energy effective Lagrangian in the $\sigma$-$\omega$ model using the Hartree approximation. 

We start with the Lagrangian of $\sigma$-$\omega$ model which is defined at some cutoff $\Lambda_0$: 
\begin{eqnarray}
L_{\Lambda_0} &=& \bar{\psi}(i \gamma_{\mu} \partial^{\mu}-M+g_\sigma
\phi-g_\omega \gamma_{\mu}V^{\mu}) \psi
+\frac{1}{2}\partial_{\mu} \phi \partial^{\mu} \phi-\frac{1}{2}m_\sigma^2
\phi^2
\nonumber \\
 & & -\frac{1}{4}F_{\mu \nu}F^{\mu \nu}
+\frac{1}{2}m_\omega^2V_{\mu}V^{\mu}+\delta L,
\nonumber \\
\delta L &=& \sum_{n=0}^4 C_n(g_\sigma \phi)^n,
\label{eq:l}
\end{eqnarray}
where $\psi$, $\phi$, $V_{\mu}$, $M$, $m_\sigma$, $m_\omega$, $g_\sigma$ and $g_\omega$ are the nucleon field, $\sigma$-meson field, $\omega$-meson field, nucleon mass, $\sigma$-meson mass, $\omega$-meson mass, $\sigma$-nucleon coupling and $\omega$-nucleon coupling, respectively. 
The vector field strength is given by $F_{\mu \nu}=\partial_{\mu}V_{\nu}-\partial_{\nu}V_{\mu}$, 
and parameters $C_n$ are constant parameters adjusted to reproduce renormalization conditions. 
In our model, the Lagrangian (\ref{eq:l}) is valid only in low-energy region below some energy scale $\Lambda_0$. 
If we consider the limit $\Lambda_0 \to \infty$, our model is equivalent to the Lagrangian in the relativistic Hartree calculation with an ordinary renormalization procedure. \cite{rf:Chin} 
However, it is natural to consider that QHD is a low-energy effective theory of QCD and is valid only below some energy scale $\Lambda_0$. 
Therefore, we keep $\Lambda_0$ finite. 
Below, we call $\Lambda_0$ a "physical cutoff" and assume that it is of the order of a few GeV. 

At present, we do not know an exact value of $\Lambda_0$.
The value should be derived from more fundamental theory QCD rather than from QHD itself. 
However, as is shown below, we do not need the exact value of $\Lambda_0$ if we are interested in the low-energy physics. 

Consider some artificial cutoff $\Lambda$ which is smaller than $\Lambda_0$. 
Of course, $\Lambda$ should be larger than the energy scale of physics in which we are interested. 
Suppose we have integrated quantum fluctuations from $\Lambda_0$ to $\Lambda$, as is in the case of the NPRG method. 
In the case of the relativistic Hartree calculation, 
this integration yields new effective terms to the Lagrangian as 
\begin{eqnarray}
L_\Lambda = L_{\Lambda_0}+\Delta L=L_{\Lambda_0}+ \sum_{n=0}^\infty D_n(g_\sigma \phi)^n,
\label{eq:el}
\end{eqnarray}
where $D_n$ are some constant parameters. 
This is an effective Lagrangian at the smaller cutoff $\Lambda$. 
This effective Lagrangian has higher ($n>4$) coupling terms of $\sigma$-meson self-interactions which were called "nonrenormalizable" in the framework of the ordinary renormalization procedure. 
However, as is seen below, they cause no difficulty. 

The parameters $D_n$ depend on $\Lambda$. 
It should be emphasized that physical quantities do not depend on the artificial  cutoff $\Lambda$, although they depend on the physical cutoff $\Lambda_0$. 
This means that the $\Lambda$-dependence of $D_n$ is canceled by the $\Lambda$-dependence of the low-energy quantum fluctuations which are calculated by using the effective Lagrangian (\ref{eq:el}) with the artificial cutoff $\Lambda$. 

If we know the exact value of $\Lambda_0$, in principle, starting with the original Lagrangian (\ref{eq:l}) at $\Lambda_0$, we can calculate $D_n$.
However, we can not do so, since we do not know the exact value of $\Lambda_0$. Alternatively, we determine $D_n$ phenomenologically. 
Suppose that $\Lambda$ is the same order of $\Lambda_0$. 
In this case, by dimensional analyses, we know that $D_n$ is at most of the order of $\Lambda^{4-n}$. 
Therefore, the higher order terms in $\Delta L$ are negligible if $\Lambda$ is larger than the energy scale of physics in which we are interested. 
In particular, higher parameters $D_n(n>4)$ vanish if we take the limit of $\Lambda\rightarrow \infty$. 
In this case, only the parameters $D_0\sim D_4$ of the "renormalizable" interactions" should be determined phenomenologically. 
Furthermore, the terms with $D_0\sim D_4$ can be included in the terms with $C_0\sim C_4$ by the redefinition. 
This is nothing but the ordinary renormalization of the renormalizable interactions. 

The physical results has $\Lambda$-dependence of the order of $\Lambda^{4-N}$, when the higher ($n\geq N$) terms in $\Delta L$ are neglected. 
Therefore, we should determine more higher terms if we want to make the $\Lambda$-dependence smaller. 
In this paper, we remove the $\Lambda$-dependence of order $\Lambda^{-2}$ in the Hartree calculation for an effective nucleon mass. \cite{rf:Kouno1,rf:Kouno2} 
This means that we use an approximated effective Lagrangian as 
\begin{eqnarray}
L_\Lambda = L_{\Lambda_0}+\Delta L=L_{\Lambda_0}+ \sum_{n=0}^6 D_n(g_\sigma \phi)^n. 
\label{eq:ela}
\end{eqnarray}
As is pointed out above, the terms with the coefficients $D_0 \sim D_4$ have been included in the terms with $C_0 \sim C_4$, 
and they are determined by the same renormalization conditions as in the ordinary relativistic Hartree approximation (RHA). \cite{rf:Chin} 
These phenomenological parameterizations of $C_0\sim C_4$ remove the $\Lambda$-dependence of the order of $\log{\Lambda }$ or of the larger order from the physical results. 
Also we must determine two coefficients $D_5$ and $D_6$ phenomenologically, 
to remove the $\Lambda$-dependence which is of the order of $\Lambda^{-2}$. 
In the references \cite{rf:Kouno1} and \cite{rf:Kouno2}, we showed that $D_5$ and $D_6$ could be determined, 
if the effective nucleon mass $M^*_0$ and the incompressibility $K$
were given in addition to the saturation conditions. 
(Here, the subscript $0$ denotes that the corresponding physical quantity 
is the one at the normal density $\rho_0$. ) 
We give the input parameters $M^*_0$, $K$, $\rho_0$ and the binding energy $E_{b0}$,and determine $C_{\sigma}=g_{\sigma}M/m_{\sigma}$, $C_{\omega}=g_{\omega}M/m_{\omega}$, $D_5$ and $D_6$. 
Note that, if $M^*_0$ is given, $C_{\omega}$ is determined 
by the relation 
\begin{eqnarray}
M_0^*=\sqrt{[M+E_{b0}-C_\omega^2 \rho_0/M^2]^2-k_{F0}^2}, 
\label{eq:h-v}
\end{eqnarray}
where $k_{F0}$ is the Fermi momentum at the normal density. 
Equation (\ref{eq:h-v}) is derived from the Hugenholtz-van Hove theorem. \cite{rf:Hugen} 
The other three parameters are determined by the saturation conditions and 
incompressibility $K$. 
We put $E_{b0}$ = $-$15.75 MeV at $\rho_0$ = 0.15 fm$^{-3}$
from the saturation conditions, and treat $M^*_0$ and $K$ as variable inputs.

Some parameter sets $M^*_0$ and $K$ are showed in Table \ref{ps}. 
We adopted $K$ = 250 $\sim$ 350 MeV 
which are consistent with empirical analyses by the relativistic nuclear models. \cite{rf:Kouno4,rf:Kouno5,rf:Kouno6} 

\vspace{0.5cm}
\begin{center}
  \begin{tabular}{c} 
     \hline table \ref{ps} \\ \hline
  \end{tabular}
\end{center}
\vspace{0.5cm}

\begin{figure}
\begin{center}
    \includegraphics*[height=8cm]{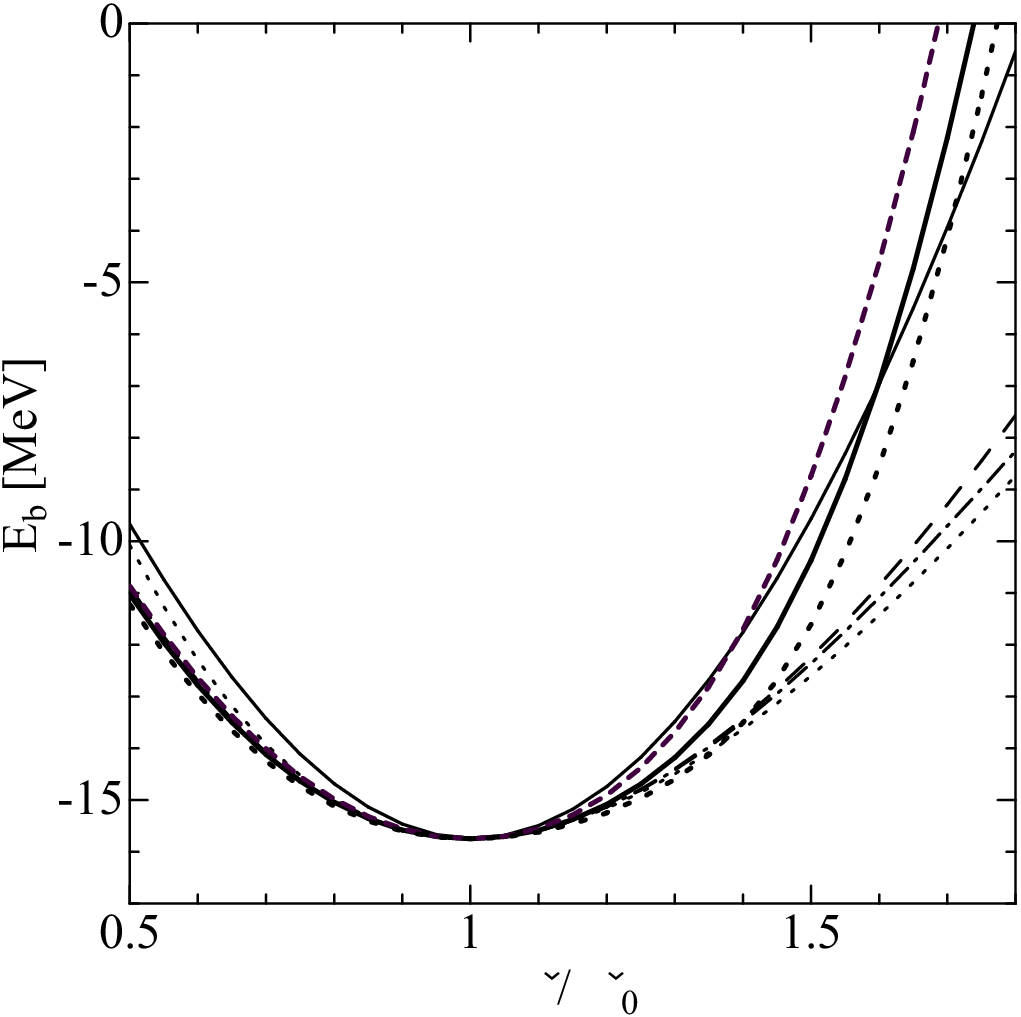}
\caption{
The binding energy Eb is shown as a function of the baryon density $\rho$.
The solid, dotted, dashed, dashed-dotted, bold solid, bold dotted and bold dashed
lines correspond to the results with the parameter sets in table \ref{ps},
RHA, PS1, PS2, PS3, PS4, PS5 and PS6 respectively.
}
\label{BE}
\end{center}
\end{figure}

\begin{figure}
\begin{center}
    \includegraphics*[height=8cm]{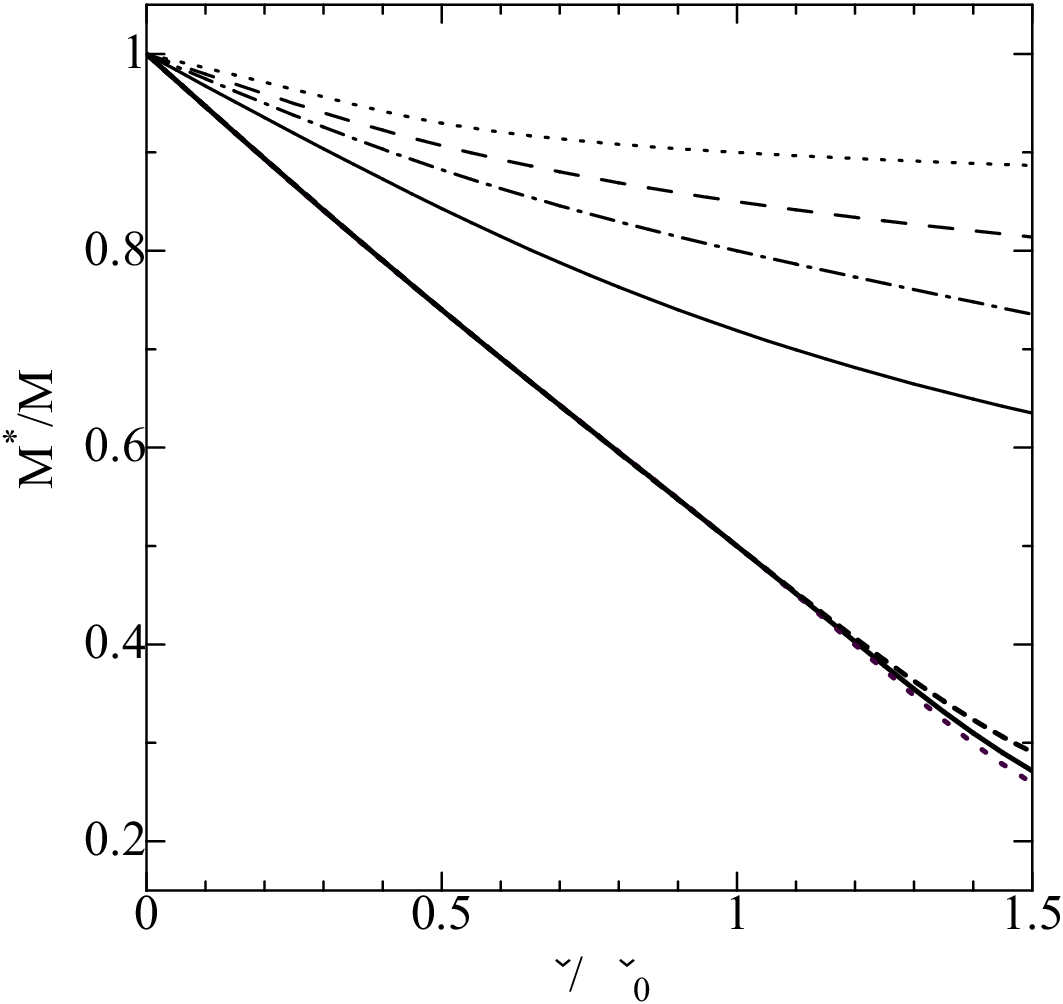}
\caption{
The effective nucleon mass $M^*$ is shown as a function
of the baryon density $\rho$.
Each line have the same correspondence to the parameter sets as
in Fig. \ref{BE}.
}
\label{M}
\end{center}
\end{figure}

In Fig. \ref{BE} and Fig. \ref{M}, we show the density dependences of the binding energy and the effective nucleon mass with the parameter sets in Table \ref{ps}. 
We see that the density dependence of $M^*$ is almost determined by $M^*_0$ and hardly depends on $K$. 

In Fig. \ref{BE} and Fig. \ref{M}, we put $\Lambda$ = 1.5 GeV.  
However, the results hardly depend on the value of $\Lambda$, since we have removed the $\Lambda$-dependence which is of the order of $\Lambda^{-2}$. \cite{rf:Kouno2} 

\vspace{0.5cm}
\begin{center}
  \begin{tabular}{c} 
     \hline Fig. \ref{BE} \\ \hline \\
     \hline Fig. \ref{M} \\ \hline
  \end{tabular}
\end{center}
\vspace{0.5cm}

\section{Vector meson selfenergies in random phase approximation}

$\quad \,$ In Sec 2, we have derived the low-energy effective Lagrangian 
in the $\sigma$-$\omega$ model using the Hartree approximation. 
Here we add the terms of $\rho$-meson, a mass of which is considered to be an important observable in the nuclear medium, to the Lagrangian. 
However, since we restrict our discussions to symmetric nuclear medium, 
this modification does not affect the results of the Hartree calculation for the effective nucleon mass and binding energy in the previous section. 

As a result, the vector meson-nucleon interaction in the Lagrangian is modified as 
\begin{eqnarray}
L_{int} &=& \bar{\psi} \Gamma^{\alpha}_{\mu} V^{\mu}_{\alpha} \psi,
\label{eq:il} \\
\Gamma^{\alpha}_{\mu} &=& g_{\alpha}
\left[\gamma_{\mu} \tau_{\alpha}-\frac{\kappa_{\alpha}}{2M}
\sigma_{\mu \lambda}iq^{\lambda}\tau^{\alpha} \right],
\end{eqnarray}
where $\alpha$ is running from 0 to 3,
$V_0$ is the $\omega$-meson field, $V_{1}\sim V_{3}$ are the $\rho$-meson 
fields, and $\tau_0$=1 and $\tau_1 \sim \tau_3$ are the isospin Pauli matrices. 
We assume that the $\omega$-nucleon tensor coupling $\kappa_{\omega}$ is negligible and put $\kappa_{\omega}=0$. 
We do not put $\kappa_{\rho}=0$. 
The numerical values of coupling constants ($g_{\rho}$, $\kappa_{\rho}$)
will be specified in the Sec.4. 

It should be emphasized that, in our framework, the interaction Lagrangian Eq. (\ref{eq:il}) is defined at finite cutoff $\Lambda$ as well as Eq. (\ref{eq:el}).  
Therefore, we assume that the tensor coupling is of order $\Lambda^{-2}$ as is shown in references. \cite{rf:Lepage,rf:Kouno3}

In the one-loop level, the RPA self-energies of the $\omega$ and $\rho$ mesons 
are given by 
\begin{eqnarray}
\Pi^{\alpha \beta}_{\mu \nu}(q) = -i\int \frac{d^{4}k}{(2 \pi)^4}
Tr[\Gamma^{\alpha}_{\mu}G(k) \tilde {\Gamma}^{\beta}_{\nu}G(k+q)],  
\label{eq:fpi}
\end{eqnarray}
where the nucleon propagator $G(k)$ in nuclear medium is given by 
\begin{eqnarray}
G(k) &=& (\gamma^{\mu}k^{\ast}_{\mu}+M^{\ast})
\left[\frac{1}{k^{\ast2}-M^{\ast2}+i\epsilon}
+\frac{i\pi}{E^{\ast}(k)} \delta(k_{0}^{\ast}
-E^{\ast}(k))\theta(k_{F}-|\mbox{\boldmath $k$}|) \right] \nonumber \\
&\equiv& G_{F}(k)+G_{D}(k),
\label{eq:g}
\end{eqnarray}
with $k^{\ast\mu} =(k^0-g_\omega V^0,\mbox{\boldmath $k$})$ and $E^{\ast}
(k) = \sqrt{\mbox{\boldmath $k$}^2+M^{\ast 2}}$. 
As for the effective nucleon mass $M^*$, we adopt results in Fig. \ref{M}.

Equation (\ref{eq:fpi}) is divided into two parts
\begin{eqnarray}
\Pi_{\mu \nu}
= \Pi^{Den}_{\mu \nu}+\Pi^{Vac}_{\mu \nu}.
\label{eq:pi1} 
\end{eqnarray}
The first term describes particle-hole excitation and the Pauli blocking for $N$ and $\bar{N}$ excitation below the Fermi surface. 
Furthermore, the density part $\Pi^{Den}_{\mu \nu}$ is divided in three parts. 
\begin{eqnarray}
\Pi^{Den}_{\mu \nu} &=& (\Pi^{Den}_v+\Pi^{Den}_{vt}+\Pi^{Den}_t)_{\mu \nu},
\\
\Pi^{Den}_v &=& \frac{g_{\alpha}^2}{4 \pi^3}
\int d^4k
\frac{1}{E^{\ast}(k)}
\delta(k_{0}^{\ast}-E^{\ast}(k))
\theta(k_{F}-|\mbox{\boldmath $k$}|)
\nonumber \\
& & \times \left[
[(w_{\mu}+k_{\mu})k_{\nu}
+(M^{*2}-wk)g_{\mu \nu}] \frac{1}{w^2-M^{*2}}
+(w \to v)
\right], \\
\Pi^{Den}_{vt} &=& - \frac{1}{8 \pi^3}
\frac{g_{\alpha}^2 \kappa_{\alpha}}{M}
\int d^4k
\frac{1}{E^{\ast}(k)}
\delta(k_{0}^{\ast}-E^{\ast}(k))
\theta(k_{F}-|\mbox{\boldmath $k$}|)
\nonumber \\
& & \times \biggl[
[+(q_{\mu}w_{\nu}+w_{\mu}q_{\nu})M^*
-(q_{\mu}k_{\nu}+k_{\mu}q_{\nu})M^*
+2(qk-wq)M^*g_{\mu \nu}]
\nonumber \\
& & \times \frac{1}{w^2-M^{*2}}
+(w \to v)
\biggr], \\
\Pi^{Den}_t &=& - \frac{1}{16 \pi^3}
\left( \frac{g_{\alpha} \kappa_{\alpha}}{M} \right)^2
\int d^4k
\frac{1}{E^{\ast}(k)}
\delta(k_{0}^{\ast}-E^{\ast}(k))
\theta(k_{F}-|\mbox{\boldmath $k$}|)
\nonumber \\
& & \times \biggl[
[(M^{*2}+wk)q_{\mu}q_{\nu}
-qk(w_{\mu}q_{\nu}+q_{\mu}w_{\nu})
-qw(q_{\mu}k_{\nu}+k_{\mu}q_{\nu})
\nonumber \\
& & +q^2(w_{\mu}k_{\nu}+k_{\mu}w_{\nu})
+\{2(qk)(wq)-(wk)q^2-q^2M^{*2} \}g_{\mu \nu}]
\nonumber \\
& & \times \frac{1}{w^2-M^{*2}}
+(w \to v)
\biggr],
\end{eqnarray}
where $v=k+q$, $w=k-q$. 
The second term in Eq. (\ref{eq:pi1}) describes $N$ and $\bar{N}$ excitation above the Fermi surface. 
This part generally has divergence and we introduce a cutoff $\Lambda$ to regularize it. 

The $\Pi^{Vac}$ calculated by using the cutoff $\Lambda$ is given by 
\begin{eqnarray}
\Pi^{Vac}_{\mu \nu}
&=& \left(\frac{q_{\mu}q_{\nu}}{q^2}-g_{\mu \nu} \right)
\Pi^{Vac} \nonumber \\
&=& \left(\frac{q_{\mu}q_{\nu}}{q^2}-g_{\mu \nu} \right)
(\Pi^{Vac}_v+\Pi^{Vac}_{vt}+\Pi^{Vac}_t), \\
\Pi^{Vac}_v &=& - \frac{g_{\alpha}^2}{2 \pi^2}
\int^1_0 dx x(1-x)q^2 F, \\
\label{eq:pi-v}
\Pi^{Vac}_{vt} &=& - \frac{1}{4 \pi^2}
\frac{g_{\alpha}^2 \kappa_{\alpha}}{M}
\int^1_0 dx q^2M^* F, \\
\label{eq:pi-vt}
\Pi^{Vac}_t &=& - \frac{1}{16 \pi^2}
\left(
\frac{g_{\alpha} \kappa_{\alpha}}{M}
\right)^2
\int^1_0 dx q^2[A^{*2}+2q^2x(1-x)] F,
\label{eq:pi-t}
\end{eqnarray}
where
\begin{eqnarray}
F &=& \log \frac{\Lambda^2+A^{*2}}{A^{*2}}
+\frac{A^{*2}}{\Lambda^2+A^{*2}}-1, \\
A^{*2} &=& M^{*2}-q^2x(1-x).
\end{eqnarray}

The $\Pi^{Vac}$  diverges if we make $\Lambda$ infinity and the renormalization is needed to remove the divergence. 
The interaction (\ref{eq:il}) is not renormalizable in the conventional sense. 
However, as is seen in the previous section, the coefficients of nonrenormalizable interaction has a suppression factor $\Lambda^{-2}$ and they cause no difficulty, even if we take the limit $\Lambda\rightarrow \infty$. 
Also remember that, in our picture, the renormalization is just removing strong $\Lambda$-dependence from the physical results. 

Because $\kappa_{\alpha}$ is of the order of $\Lambda^{-2}$, direct calculations show that the $\Lambda$-dependence of $\Pi^{Vac}_{v}$, $\Pi^{Vac}_{vt}$ and $\Pi^{Vac}_{t}$
are of the order of $\log \Lambda$, $(\log \Lambda)/\Lambda^2$ and $(\log \Lambda)/\Lambda^4$
respectively. 
As is in the case of the ordinary renormalization procedure, 
we expand $\Pi^{Vac}$ around zero density $M^*=M$ and
on-shell momentum $q^2=m_v^2$,
\begin{eqnarray}
\Pi^{Vac}(q,M^*,\Lambda)
&=& \sum_{l=0}^{\infty} \sum_{m=0}^{\infty}
G_{l,m}(q^2-m_v^2)^l(M^*-M)^m
\nonumber \\
&=& \sum_{n=0}^{\infty} \frac{1}{n!}
\left[
(q^2-m_v^2) \frac{\partial}{\partial q^2}
+(M^*-M) \frac{\partial}{\partial M^*}
\right]^n
\Pi^{Vac}(q,M^*,\Lambda) \bigg|_{{M^*=M \atop q^2=m_v^2}},
\label{eq:ex}
\end{eqnarray}
where $m_v$ is the $\omega$ or the $\rho$ mass at zero density and $G_{l,m}$'s are some parameters which depend on $\Lambda$. 
Direct calculations show that the coefficients 
$G_{0,0}$ and $G_{1,0}$ have the $\Lambda$-dependence of the order of $\log \Lambda$, 
the coefficients $G_{0,1}$ and $G_{1,1}$ have the $\Lambda$-dependence of the order of $(\log \Lambda)/\Lambda^2$, and the coefficients of the other higher terms have the $\Lambda$-dependence which is smaller than $O(\log{\Lambda}/\Lambda^2)$. 
We also remark that the coefficients $G_{0,1}$ and $G_{1,1}$ have originated in the vector-tensor part $\Pi_{vt}^{Vac}$. 

In the RPA meson self-energy, we remove the $\Lambda$-dependence which is $O(\log{\Lambda}/\Lambda^{2})$ or larger. 
As is in the previous section, this is achieved by adding some additional terms 
\begin{eqnarray}
\delta L'=G_{0,0}'V_{\mu}V^{\mu} +G_{1,0}'F_{\mu \nu}F^{\mu \nu}
+G_{0,1}'\phi V_{\mu}V^{\mu} +G_{1,1}'\phi F_{\mu \nu}F^{\mu \nu}.
\label{eq:el2}
\end{eqnarray}
to the original Lagrangian. 
If we determine the coefficients $G_{0,0}'$ and $G_{1,0}'$ phenomenologically, we can remove $\Lambda$-dependence of $O (\log \Lambda)$ of the meson self-energies. 
The term with $G_{0,0}'$ is vector-meson mass counterterm and should cancel the term with $G_{0,0}$. 
The term with $G_{1,0}'$ is vector-meson wave function counterterm and should cancel the term with $G_{1,0}$. 
Therefore, we get $G_{0,0}'=-G_{0,0}$ and $G_{1,0}'=-G_{1,0}$. 
These two counterterms should be needed in the case of ordinary renormalization procedure, i.e., in the limit $\Lambda \to \infty$. 
Moreover, we can remove the errors of the order of $(\log \Lambda)/\Lambda^2$
by determining $G_{0,1}'$ and $G_{1,1}'$ phenomenologically. 

Since we put $\kappa_{\omega}=0$, $\Pi_{vt}$ = $\Pi_t$ = 0 in the $\omega$-meson self-energy, 
the coefficients $G_{\omega 0,1}$ and $G_{\omega 1,1}$ are zero, and, therefore, $G_{\omega 0,1}'$ and $G_{\omega 1,1}'$ are not needed. 
As a result, the $N \bar{N}$ contribution in the $\omega$-meson self-energy is given by 
\begin{eqnarray}
\Pi^{Vac}_{re}
= \Pi^{Vac}
-G_{\omega 0,0}-G_{\omega 1,0}(q^2-m_\omega^2).
\label{eq:re-pi}
\end{eqnarray}
On the other hand, in the $\rho$-meson self-energy, 
we must determine the coefficients $G_{\rho 0,1}'$ and $G_{\rho 1,1}'$ 
to remove the $\Lambda$-dependence of the order of $(\log{\Lambda})/\Lambda^2$. 
The $N \bar{N}$ contribution of the $\rho$-meson self-energy is given by 
\begin{eqnarray}
\Pi^{Vac}_{re}
= \Pi^{Vac}
-G_{\rho 0,0}-G_{\rho 1,0}(q^2-m_\rho^2)
+G_{\rho 0,1}'(M^*-M)
+G_{\rho 1,0}'(q^2-m_\rho^2)(M^*-M).
\label{eq:re-pi}
\end{eqnarray}
In the next section, we determine the coefficients $G_{\rho 0,1}'$ and $G_{\rho 1,1}'$ to reproduce the results obtained by the QCD sum rule. \cite{rf:Hatsuda}

\section{Effective vector meson masses in nuclear medium}

$\quad \,$ The effective meson mass is defined as the pole of 
the meson propagator in the nuclear medium, namely, given by the following equation, 
\begin{eqnarray}
m_v^{* 2}=m_v^2+\Pi_L(q_0=m_v^*,{\mbox{\boldmath $q$}}=0), 
\label{eq:propa}
\end{eqnarray}
where $\Pi_L$ is longitudinal component of meson self-energy $\Pi_{\mu\nu}$.

In this section, at first, we put $\Lambda =$1.5GeV. 
The $\Lambda$-dependence of our results will be checked at the end of this section. 

In Fig. \ref{m-M}, 
we show the $M^*_0$-dependence of the effective $\omega$-meson mass $m_{\omega 0}^*$ at the normal density. 
The $m_{\omega 0}^*$ decreases as $M^*_0$ decreases. 
It should be emphasized that this result does not depend on the value of the incompressibility $K$. 
In the RPA calculation of $\omega$-meson self-energy, only $g_v$ and $M^*$ are variable inputs. 
Furthermore, at the normal density, $g_v$ is uniquely determined via Eq. (\ref{eq:h-v}), if $M^*_0$ is given. 
Therefore, the effective $\omega$-meson mass is also uniquely determined at the normal density, if $M^*_0$ is given. 
As $M_0^*$ becomes smaller, $g_{\omega}$ becomes larger via Eq. (\ref{eq:h-v}).
As a result, the effective $\omega$-meson mass becomes smaller, as $M_0^*$ becomes smaller. 

\begin{figure}
\begin{center}
    \includegraphics*[height=8cm]{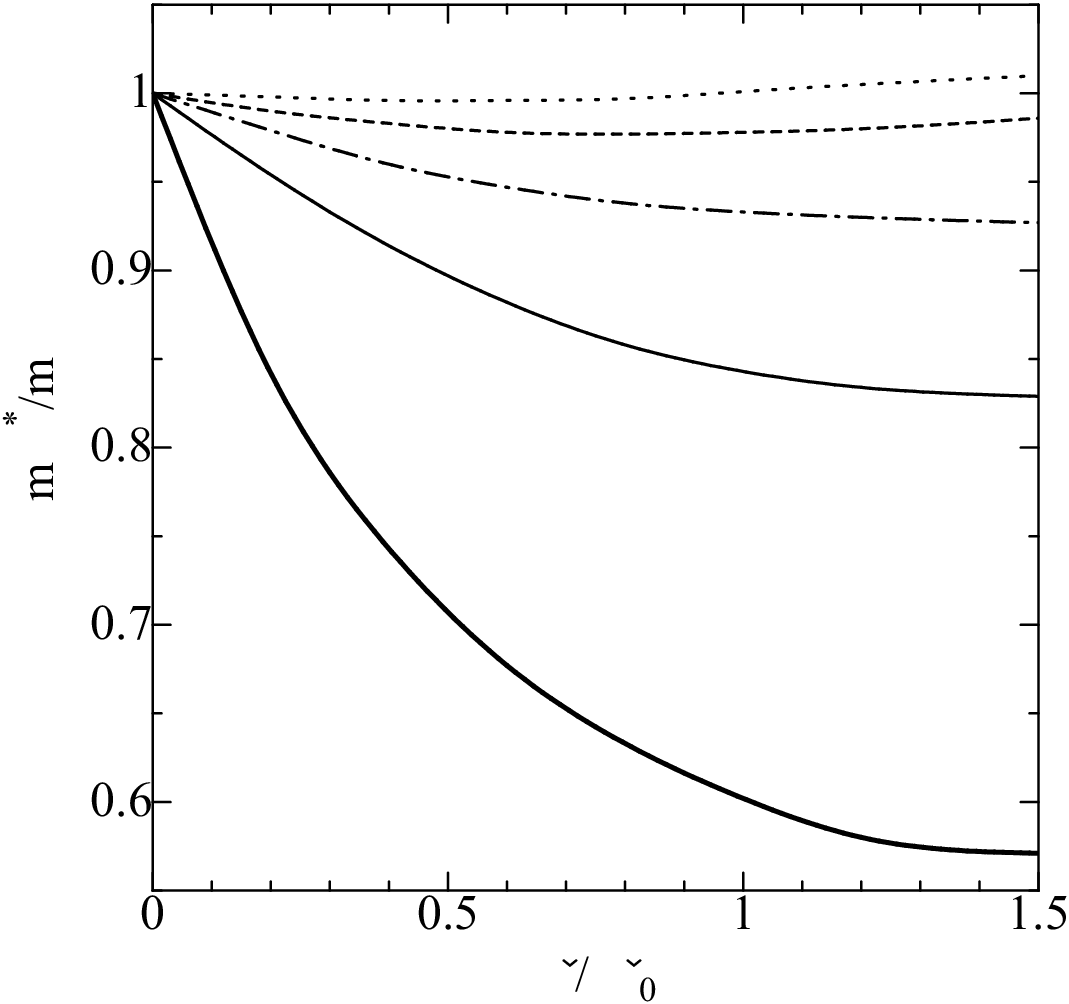}
\caption{
The effective $\omega$-meson mass $m_{\omega}^*$ is shown
as a function of the baryon density $\rho$.
Each line have the same correspondence to the parameter sets as
in Fig. \ref{BE}.
}
\label{mo}
\end{center}
\end{figure}

In Fig. \ref{mo}, we show the density dependence of $m_\omega^*$ with parameter sets PS1 $\sim$ PS4 in table \ref{ps}. 
In the cases of PS3 and PS4, $m_\omega^*$ decreases as $\rho$ increases. 
However, decreases are much smaller than in the result obtained by using the ordinary RHA calculation for $M^*$. 
In the cases of PS1 and PS2, $m_\omega^*$ hardly decreases. 
In the parameter sets PS1$\sim$PS4, we put $K=300$MeV. 
However, as shown in Fig. \ref{M}, the $K$-dependence of $M^*$ is very small. 
Therefore, the $K$-dependence of $m_\omega^*$ is also very small, since $m_\omega^*$ depends on $K$ only via $M^*$.

\vspace{0.5cm}
\begin{center}
  \begin{tabular}{c} 
     \hline Fig. \ref{m-M} \\ \hline \\
     \hline Fig. \ref{mo} \\ \hline
  \end{tabular}
\end{center}
\vspace{0.5cm}

\begin{figure}
\begin{center}
    \includegraphics*[height=8cm]{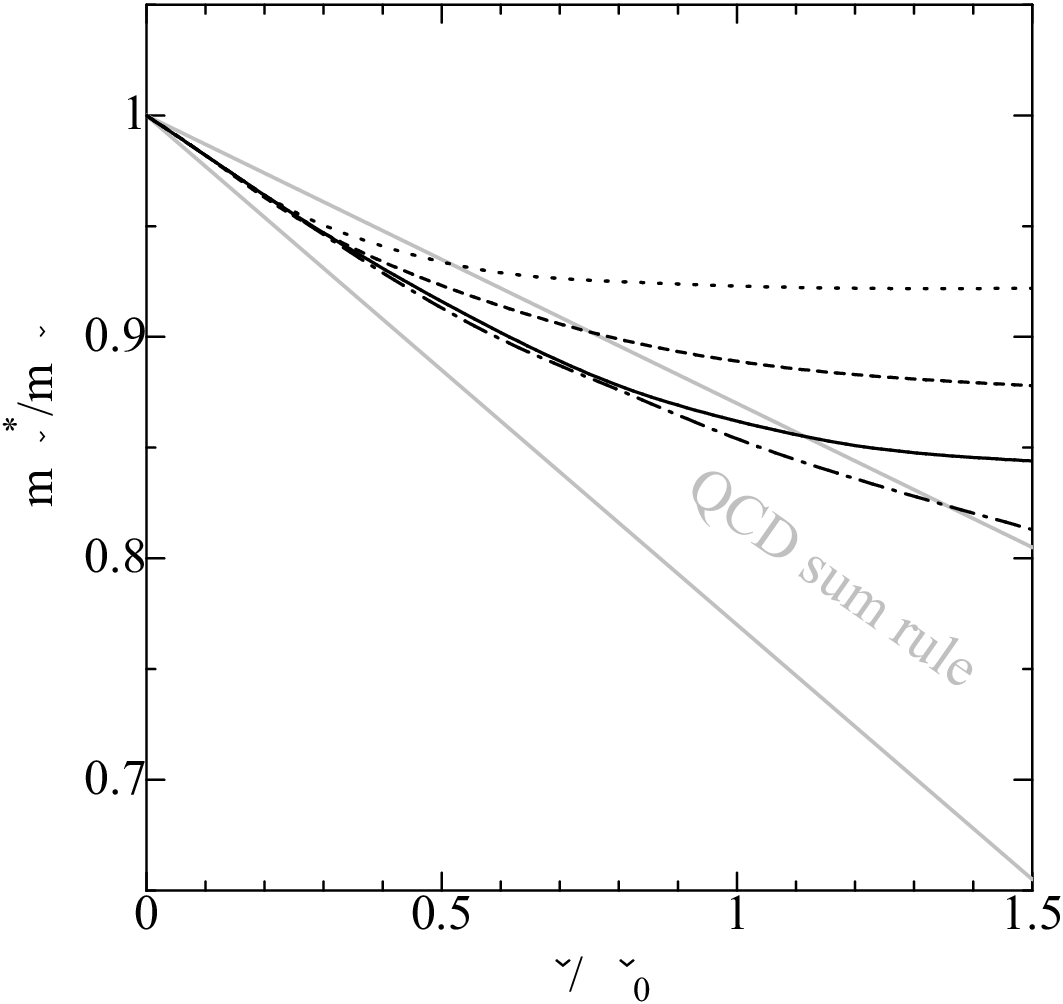}
\caption{
The effective $\rho$-meson mass $m_{\rho}^*$ is shown
as a function of the baryon density $\rho$.
Each line have the same correspondence to the parameter sets as
in Fig. \ref{BE}.
}
\label{mr}
\end{center}
\end{figure}

In Fig. \ref{mr}, we show the density dependence of the effective $\rho$-meson mass in nuclear medium. 
In the numerical calculations, we have put ($g_{\rho}$,$\kappa_{\rho}$) = (2.63,6.0) which are obtained from the $N$-$N$ forward dispersion relation in Ref. \cite{rf:Grein}. 
Moreover, we use the following linear relation obtained by the QCD sum rule in Ref. \cite{rf:Hatsuda} as the phenomenological inputs. 
\begin{eqnarray}
\frac{m_{\rho QSR}^*}{m_{\rho}}=1-(0.18 \pm 0.05) \frac{\rho}{\rho_0}
\label{eq:S-H}
\end{eqnarray}
It is considered that the linear relation (\ref{eq:S-H}) works well at low density. 
Therefore, we have determined the coefficients $G_{\rho 0,1}'$ and $G_{\rho 1,1}'$ to reproduce the mean value of the linear relation (\ref{eq:S-H}) at low density($\rho/\rho_0<0.2$). 

In Fig \ref{mr}, as the density increases, the $m^*_{\rho}$ decreases. 
At higher density, the rate of the decrease becomes smaller. 
As a result, the $m^*_\rho/m_\rho$ is 0.85$\sim$0.95 at the normal density. 

Comparing Fig. \ref{mr} and Fig. \ref{M}, we see that there is a strong correlation between $m_\rho^*$ and $M^*$ in the CFT calculations. 
At high density, the $m_\rho^*$ becomes smaller as $M^*$ becomes smaller.

\vspace{0.5cm}
\begin{center}
  \begin{tabular}{c} 
     \hline Fig. \ref{mr} \\ \hline
  \end{tabular}
\end{center}
\vspace{0.5cm}

\begin{figure}
\begin{center}
    \includegraphics*[height=8cm]{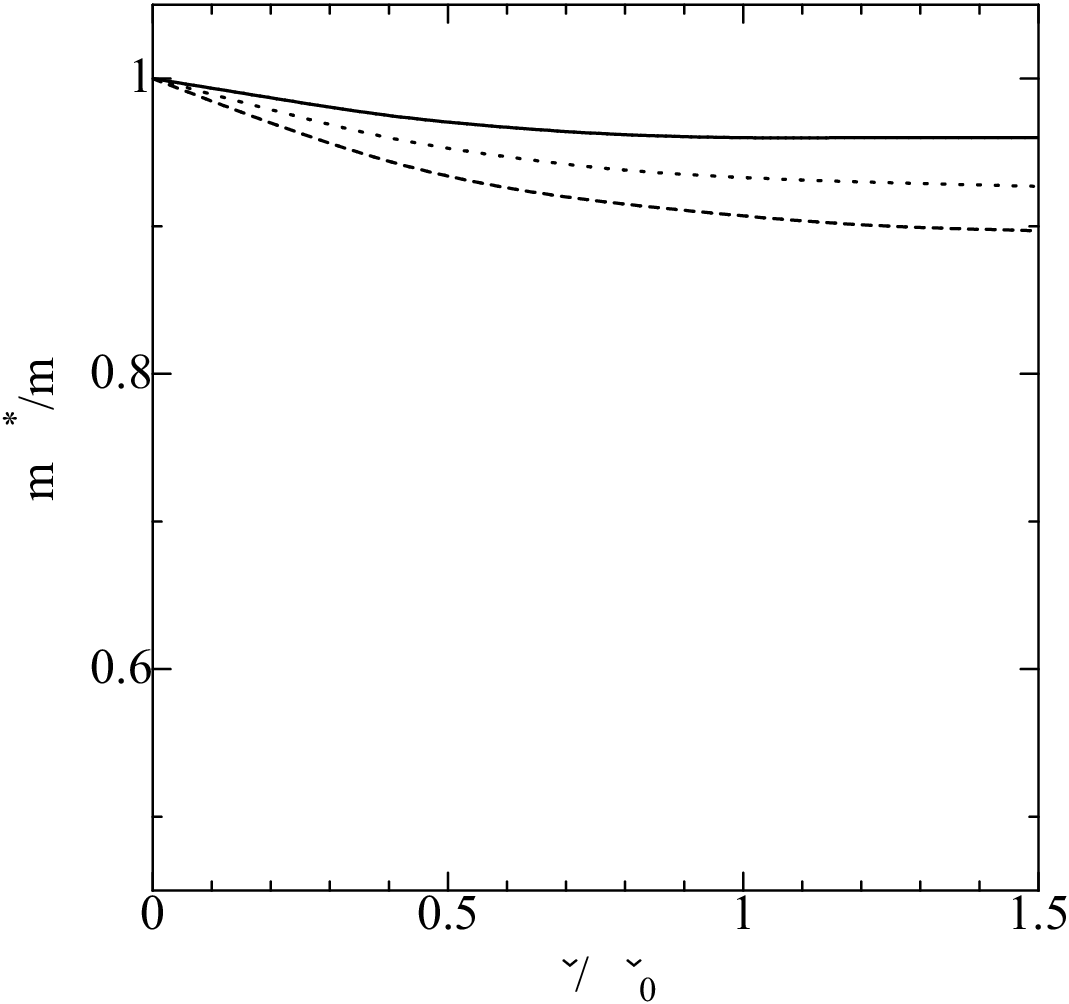}
\caption{
The $\Lambda$-dependence of the effective 
$\omega$-meson mass with the parameter set PS3.
The solid, dotted and dashed
lines correspond to the results with the cutoff $\Lambda$ =1.2, 1.5 and 2
GeV respectively.
}
\label{mo-c}
\end{center}
\end{figure}

\begin{figure}
\begin{center}
    \includegraphics*[height=8cm]{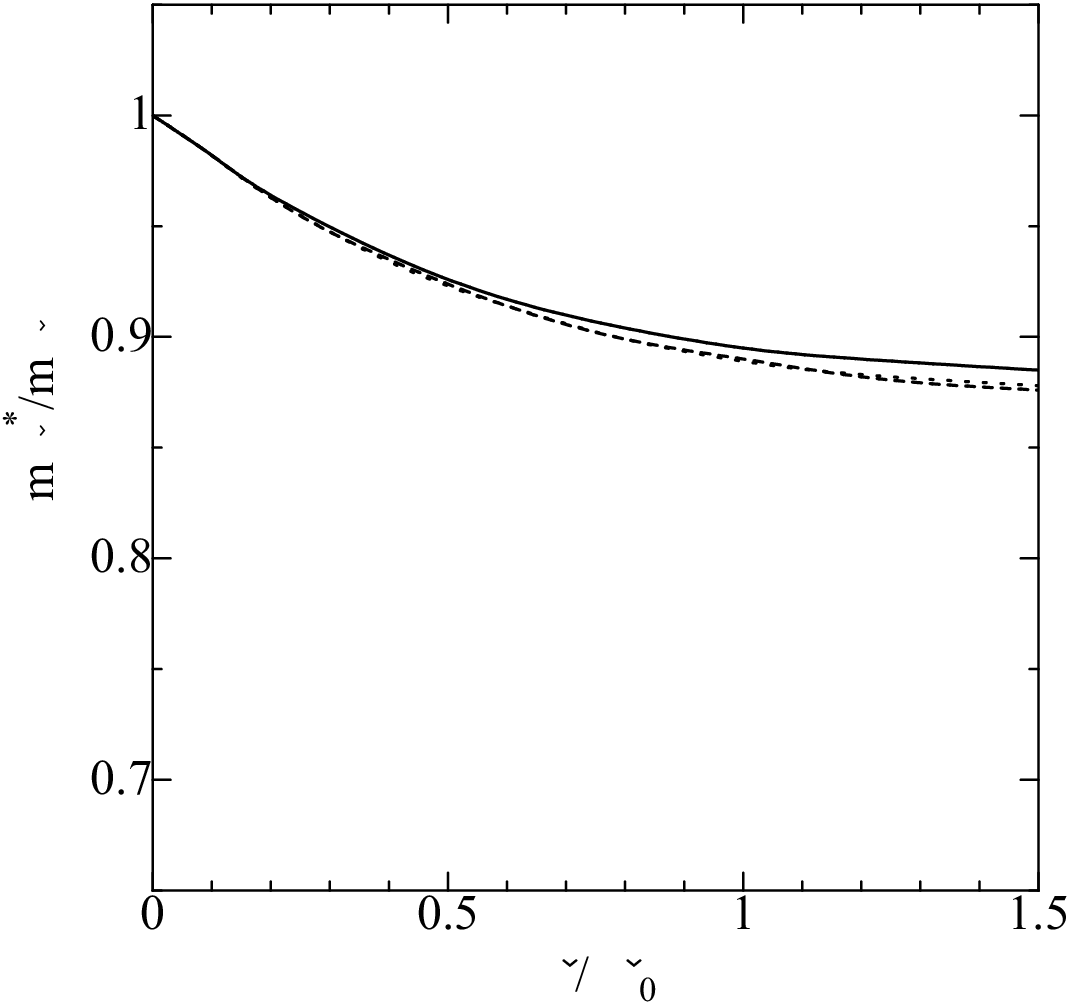}
\caption{
The $\Lambda$-dependence of the effective
$\rho$-meson mass with the parameter set PS2.
Each line have the same correspondence to the parameter sets as
in Fig. \ref{mo-c}.
}
\label{mr-c}
\end{center}
\end{figure}

Finally, we examine the $\Lambda$-dependence of our results. 
In Fig. \ref{mo-c} and Fig. \ref{mr-c}, we show  the $\Lambda$-dependence of
effective masses of $\omega$ and $\rho$-mesons. 
In our framework, the tensor coupling $\kappa$ should have a $\Lambda$-dependence. 
For the $\rho$-$N$ coupling, we assume 
\begin{eqnarray}
\kappa_{\rho}={c\over{\Lambda^2}}, 
\label{eq:kappa}
\end{eqnarray}
where $c$ is a constant which does not depend on $\Lambda$. 
Putting $\kappa_\rho=6.0$ and $\Lambda=1.5$GeV, we get $c=13.5$GeV$^{2}$. 
For the $\omega$-meson, we neglect the tensor coupling $\kappa_{\omega}$ for any value of $\Lambda$. 
Although we remove the $\Lambda$-dependence only of the order $(\log{\Lambda})/\Lambda^2$ and of the larger order, the $\Lambda$-dependence is less than 10 percents of the effective vector meson masses itself. 
In particular, $m_\rho^*$ hardly depends on $\Lambda$, 
since we determine $G_{\rho 0,1}'$ and $G_{\rho 1,1}'$ phenomenologically.

\vspace{0.5cm}
\begin{center}
  \begin{tabular}{c} 
     \hline Fig. \ref{mo-c} \\ \hline \\
     \hline Fig. \ref{mr-c} \\ \hline \\
  \end{tabular}
\end{center}
\vspace{0.5cm}

\section{Summary}

$\quad \,$ The results obtained in this paper are summarized as follows. 

(1) Using a low-energy effective Lagrangian, we formulated the method of renormalization for the RPA vacuum polarization effects in QHD with a finite cutoff. 
Using this formulation, we have studied the effective masses of vector mesons in nuclear medium. 

(2) At the normal density, the effective $\omega$-meson mass $m^*_{\rho0}$ is uniquely determined if the effective nucleon mass $M^*_0$ is fixed. 

(3) If $M_0^*$ is not much large ($\leq 0.80M$), $m^*_\omega$ decreases as the density increases. 

(4) To calculate the effective $\rho$-meson mass $m^*_\rho$, at low density, we fit two phenomenological parameters to reproduce the linear relation, which is obtained by the QCD sum rule. 
By these phenomenological parameterizations, 
we could remove the errors of order $(\log \Lambda)/\Lambda^2$ from our results. 

(5) The effective $\rho$-meson mass $m_\rho^*$ decreases as density increases. 
The rate of the decrease becomes smaller at higher density. 
As a result, the $m^*_\rho/m_\rho$ is 0.85$\sim$0.95 at the normal density.

(6) There is a strong correlation between $m_\rho^*$ and $M^*$ in the CFT calculations. 
At high density, the $m_\rho^*$ becomes smaller as $M^*$ becomes smaller. 

(7) The $\Lambda$-dependence of our results is checked and it is found to be small. 

In the RHA and RPA calculations, the quantum fluctuations are estimated by the one loop approximation. 
The NPRG analyses show the possibility that this approximation may work well at low energy. \cite{rf:Clark,rf:Kouno7} 
However, even if the higher-loop corrections are important, we are able to estimate them by solving the NPRG equations directly. 
These studies are in progress.


\begin{center}
{\large \bf Acknowledgements}
\end{center}
$\quad \,$ The authors thank T. Kohmura, M. Matsuzaki, T. Maruyama and L. Liu for useful discussions. 

\vspace{2cm}


\newpage


\begin{table}
\begin{center}
  \begin{tabular}{lcccccc} \hline \hline
  Model & $M_0^*/M$ & $K$ [MeV] & $C_\omega^2$ & $C_\sigma^2$ &
  $D_5M$ & $D_6M^2$ \\ \hline
  RHA & 0.72 & 462.7 & 147.84 & 228.16 & 0.0 & 0.0 \\ 
  CFT PS1 & 0.90 & 300.0 & 30.70 & 103.09 & -1.81 & 21.41 \\
  CFT PS2 & 0.85 & 300.0 & 65.15 & 149.20 & 0.17 & -0.22 \\
  CFT PS3 & 0.80 & 300.0 & 99.41 & 184.29 & 6.19 & -0.13 \\
  CFT PS4 & 0.50 & 300.0 & 298.32 & 389.02 & 7.83 & -1.67 \\
  CFT PS5 & 0.50 & 250.0 & 298.32 & 390.03 & 8.92 & -1.81 \\
  CFT PS6 & 0.50 & 350.0 & 298.32 & 388.24 & 7.03 & -1.53 \\
  \hline \hline
  \end{tabular}
\end{center}
\caption{The parameter sets in $\sigma$-$\omega$ model.}
\label{ps}
\end{table}


\begin{figure}
\begin{center}
    \includegraphics*[height=8cm]{m-M.eps}
\caption{
$m_{\omega 0}^*$-$M_0^*$ relation at the normal density $\rho_0$.
}
\label{m-M}
\end{center}
\end{figure}

\end{document}